\documentclass[%twocolumn,
aps,nofootinbib,showpacs,showkeys,preprint
tightenlines,preprintnumbers,] {revtex4}

\usepackage{epsf,epsfig,subfigure,graphicx,amsmath,amssymb}
\usepackage{color}

\newcommand{\etal}{{\it et al.}}

\newcommand{\adec}{\alpha^{\rm dec}}

\begin{document}
\draft

\title{\Large\bf New confining force solution of QCD axion domain wall problem}

\author{S. M. Barr}
\address{Department of                                                            Physics and Astronomy, Bartol Research Institute, University of Delaware, Newark, DE 19716, USA}
\author{Jihn E. Kim}
\address{Department of Physics, Kyung Hee University, 26 Gyeongheedaero, Dongdaemun-Gu, Seoul 130-701, Korea}

\begin{abstract}
The serious cosmological problems created by the axion-string/axion-domain-wall system in standard axion models are alleviated by positing the existence of a new confining force. The instantons of this force can generate an axion potential that erases the axion strings long before QCD effects become important, thus preventing QCD-generated axion walls from ever appearing. Axion walls generated by the new confining force would decay so early as not to contribute significantly to the energy in axion dark matter.

\end{abstract}


\maketitle

%\section{Introduction}
\noindent{\it Introduction}--
The idea of axion fields \cite{axions} arose in the context of the Peccei-Quinn (PQ) mechanism \cite{PQ} for solving the ``Strong CP Problem" \cite{StrongCP}, i.e. explaining why QCD interactions approximately conserve CP. It was soon realized that axion fields can have many interesting cosmological effects, including coherent axion oscillations as dark matter \cite{coherent-osc}, and topological defects in the axion field, specifically ``axion strings" and ``axion domain walls", that form in the early universe. If such defects persist they can ``overclose" the universe (i.e. dominate the present energy density in the universe) \cite{axion-domain-wall-problem}, while if they disappeared in the early universe by radiating away their energy (predominantly into axion particles) they could significantly increase the axion dark matter density, thus tightening the constraints on axion models \cite{harari-sikivie, hagmann-chang-sikivie, hiramatsu, davis, battye-shellard}.

In this letter we address the last point. We propose a mechanism that can greatly suppress the contribution to dark matter from axions radiated by strings and domain walls, and  alleviate the constraints on axion dark matter models.

In a typical axion model one has a complex scalar field, $\Phi$, with a zero-temperature potential of the ``Mexican hat" form:
$\frac{1}{4} \lambda \left[ | \Phi |^2 - (F_a/2)^2\right]^2$.
This form is dictated by the Peccei-Quinn symmetry, which is a global $U(1)$ under which the phase of $\Phi$ rotates. When the cosmic temperature falls well below $F_a$, $\Phi$ develops
a vacuum expectation value (VEV): $\Phi(x^{\mu}) = (F_a/\sqrt2) e^{i a(x^{\mu})/F_a}$. The value of  $a(x^{\mu})$ varies randomly in space, so that cosmic strings form, around which $a(x^{\mu})/F_a$
winds by $2 \pi$. As the universe expands, these strings quickly form a
scale-invariant distribution on super-horizon scales, with the number density of strings of length $\ell$ given by $n(\ell) d \ell \propto \ell^{-4} d \ell$ for $\ell > H^{-1}$, where $H$ is the Hubble parameter. String loops with $\ell < H^{-1}$ quickly radiate away their energy. Thus the energy density in strings goes as $\int^{\infty}_{H^{-1}} (F_a^2 \ell) n(\ell)
d \ell \sim F_a^2 t^{-2} \sim (F_a/M_P)^2 \rho_{total}$, which is negligible, as generally it is assumed $F_a \ll M_P$. However, as explained below, string evolution changes when $T$ reaches the QCD scale.

The PQ symmetry has a QCD anomaly, which means that at the quantum level it is violated by non-perturbative QCD effects (instantons). When the cosmic temperature falls to near the QCD scale ($\Lambda_{QCD} \sim 200$ GeV), the effects of QCD instantons ``turn on" and lift the degeneracy of the axion field, giving it a potential of the
cosine function form \cite{StrongCP},

\begin{equation}
V(a) = c \Lambda_{QCD}^4 \left(1 - \cos \left( N \frac{a}{F_a} + \theta_{QCD} \right) \right) = c \Lambda_{QCD}^4 \left(1 - \cos \left( \frac{a}{f_a} + \theta_{QCD} \right) \right) ,
\end{equation}
\noindent where $c$ is a dimensionless quantity; $N$ is a model-dependent integer that gives the strength of the QCD anomaly; and the angle $\theta_{QCD}$ is a parameter appearing in the only CP-violating term in the QCD lagrangian. Note that since $\Phi(x^{\mu}) = (F_a/\sqrt2) e^{i a(x^{\mu})/F_a}$ the field $a(x^{\mu})/F_a$ can take physically distinct values in the interval $[0, 2 \pi)$.
The spontaneously broken PQ symmetry with the shift symmetry of axion, $a\to a+{\rm constant}$, generates a non-vanishing potential proportional to $-(\Lambda_{QCD}^4\Phi+{\rm h.c.})$, where we can treat $\Lambda^4_{QCD}\to  e^{-i a(x^{\mu})/F_a}\Lambda^4_{QCD}$ under the PQ transformation. This kind of phase introduction was shown most succinctly in \cite{VafaWitten84}.
Therefore the potential in Eq. (1) has $N$ degenerate minima at $a/F_a = - \theta_{QCD}/N + 2 \pi k/N$, with $k = 0, 1, 2, ..., N-1$.  The axion field will settle 
into one of these minima, thus canceling out $\theta_{QCD}$. This is how axion models solve the Strong CP problem. It is sometimes convenient to express various quantities 
in terms of $f_a \equiv F_a/N$ so that $N$ drops out of the expressions, as we have 
done in the second equality in Eq. (1). For example, the axion mass is $c^{1/2} \Lambda_{QCD}^2/f_a$. \footnote{The distinction between $F_a$ and $f_a$ is equivalent to that after integrating out all the quarks $f_a$ is defined from the coefficient of the gluon anomaly term as $(g_s^2 a/32\pi^2 f_a)G\tilde G$ while $F_a$ is defined from the phase(s) of the PQ-charged scalar fields.}

When the QCD-instanton potential for the axions ``turns on",
the axion field typically finds itself misaligned from a minimum and starts to oscillate coherently about it \cite{coherent-osc}. These long-wavelength oscillations are equivalent to a gas of low-momentum axions, and play the role of cold dark matter. This is the ``misalignment" mechanism for producing axion dark matter. It was shown in \cite{coherent-osc} that the present energy density in these coherent axion oscillations depends linearly
on $f_a$. This gives an upper bound on $f_a$ of roughly $10^{11}$ GeV.

If $N \neq 1$ the axion field has several degenerate minima
$a/F_a = \theta_{QCD}/N + 2 k \pi/N$, $ k = 0, 1 ,2 ,\cdots, N-1$.
The regions of the universe where the axion (randomly) chooses different minima are separated by domain walls. As one goes around a string, $a/F_a$ winds by $2 \pi$, so that the axion field passes through $N$ different minima, implying that each string has $N$ walls attached to it. Thus,
once $V(a)$ turns on, a foam-like ``strings-wall network" forms that cannot disappear by radiation. Its energy density
scales as $a^{-1}$, where $a$ is the cosmic scale factor. Since energy density in relativistic and non-relativistic particles scale as $a^{-4}$ and $a^{-3}$, respectively, the string-wall system would eventually dominate by many orders of magnitude the energy in baryons and other particles. This is the famous ``axion domain wall problem" \cite{axion-domain-wall-problem}.

This disaster could be avoided if the spontaneous breaking of the PQ $U(1)$ symmetry happened before or during inflation. Then the density of strings (and thus of the walls that eventually attach to them) could be so diluted by inflation as to be harmless. However, recent analyses based on BICEP2 data suggest that this possibility is excluded \cite{visinelli-gondolo, marsh}.
The resulting isocurvature
fluctuations would have large amplitude in conflict with measurements
of the CMB temperature power  \cite{Plan13,WMAP9}.
Regarding the determination of $r$, the foreground subtraction must be carefully carried out \cite{foregrounds}.

That seems to leave only one way to avoid axion strings and walls ``overclosing" the universe, and that is to assume that $N=1$ \cite{KimDW14}. If $N=1$, walls would still form, but with every string having just one wall attached to it. As pointed out by Vilenkin and Everett \cite{vilenkin-everett}, if $N = 1$, then once the walls form they
quickly get chopped up into finite sized areas bounded by closed strings, and are able to shrink and disappear by radiating away their energy. The energy in these finite patches of wall is easily shown to be roughly comparable to the energy in the coherent axion oscillations (and much larger than that in strings). As these patches oscillate they
radiate their energy predominantly into axion particles. The ratio of the energy in the axions radiated by decaying walls to the energy in the coherent axion oscillations is denoted by $\adec$. The factor $\adec $ is the subject of a long-running controversy. $\adec $ has been calculated by different groups with widely varying results. \cite{harari-sikivie, hagmann-chang-sikivie} obtained $\sim 0.19$; \cite{hiramatsu} obtained $6.9 \pm 3.5$; and \cite{davis, battye-shellard} obtained $\sim 186$.

This highly uncertain factor $\adec $ enters in an important way into bounds (coming from the observed dark matter abundance) on the axion decay constant $f_a$ and the axion mass $m_a$. Visinelli and Gondolo \cite{visinelli-gondolo} have recently found these to be

\begin{equation}
f_a = (8.7 \pm 0.2) \times 10^{10} {\rm GeV} (\adec + 1)^{-6/7}, \;\;\;\;
m_a = (71 \pm 2) \mu{\rm eV} (\adec  + 1)^{6/7},
\end{equation}

\noindent if axions make up all the dark matter in the universe. Since there may be other significant contributions to the dark matter density, these would be an upper bound for $f_a$ and a lower bound for $m_a$.

%%%%%%%%%%%%%%%%%%%%
%\section{Elimination of strings and walls by new confining force}
\noindent{\it Elimination of strings and walls by new confining force}--
We now propose a mechanism that would greatly suppress the contribution of decaying axion strings and walls to the dark matter density. Suppose we introduce a new non-abelian gauge interaction based on the group $G_h$. Let there be a fermion $Q$ that transforms only under $G_h$, which will be called the $G_h$-quark. The Standard Model quarks transform only under the Standard Model gauge group and will be denoted by $u_i, d_i$, with $i = 1,2,3$. As typical of axion models, suppose there are two Higgs doublets, $\phi_1$ and $\phi_2$, with the up-type quarks $u_i$ coupling to $\phi_1$ and the down-type quarks $d_i$ coupling to $\phi_2$. Suppose that the field we called $\Phi$ couples to $Q$ and also to the doublet Higgs fields $\phi_1$ and $\phi_2$ (so that the global symmetry of the lagrangian that rotates the phase of $\Phi$ also rotates the relative phases of $\phi_1$ and $\phi_2$). In particular, suppose we have the following couplings:

\begin{equation}
\sum_{i=1}^3 Y_u^i \overline{u}_{Ri} u_{Li} \phi_1 + \sum_{i=1}^3 Y_d^i \overline{d}_{Ri} d_{Li} \phi_2^* + Y_Q \overline{Q}_R Q_L \Phi + g \phi_1^* \phi_2 \Phi + H.c.
\end{equation}

\noindent Thus both ordinary quarks and $Q$ are charged under the Peccei-Quinn symmetry, which therefore has both a QCD and a $G_h$ anomaly. As before, let us write  $\langle \Phi \rangle = (f/\sqrt2) e^{i a/f}$. If $f \gg v_1, v_2$, where $v_i
\equiv \langle \phi_i \rangle$, then we may write $F_a = \sqrt{ f^2 + v_1^2 + v_2^2 } \cong f$. And the zero-temperature potential for the axion field can be written

\begin{equation}
V(a) = V_{QCD}(a)  + V_h(a) = c \Lambda_{QCD}^4 \left( 1 - \cos \left( N_{QCD} \frac{a}{F_a} + \theta_{QCD} \right) \right) + c' \Lambda_h^4 \left( 1 - \cos \left( N_h \frac{a}{F_a} + \theta_h \right) \right).\label{eq:Vaxion}
\end{equation}

\noindent Here $\Lambda_h$ is the confinement scale of the $G_h$ force, which we assume $\gg \Lambda_{QCD}$; $\theta_h$ is the CP-violating phase of the $G_h$ interactions; and $N_{QCD}$ and $N_h$ are
integers giving the strength of the QCD and $G_h$ anomalies of the PQ $U(1)$ symmetry. In this model, $N_{QCD} = 6$ and $N_h = 1$.  Note that it is trivial to have $N_h = 1$, as it is not connected with the number of Standard-Model quarks.

When $T$ falls below $F_a$, strings form.
When $T \sim \Lambda_h \gg \Lambda_{QCD}$, the $G_h$ instanton effects ``turn on", generating $V_h(a)$, but $V_{QCD}(a)$ has not yet turned on. $V_h (a)$ causes axion walls to form. Since $N_h = 1$, each string has only one wall attached to it, and the Vilenkin-Everett mechanism will operate, chopping the walls into finite patches that radiate away their energy into axions. As will be seen, these ``decay axions" contribute negligibly to dark matter.

After the strings and the walls generated by $V_h (a)$ have disappeared, the axion field will settle to the value
$a/F_a = - \theta_h$, with fluctuations of order $T^2/\Lambda_h^2$. Thus, the axion field is everywhere aligned.
Consequently, when $T \sim \Lambda_{QCD}$ and $V_{QCD}$ turns on, no walls are formed (except for rare closed surfaces produced by thermal fluctuations).

One problem that is immediately apparent is that $V_h (a)$ would destroy the solution to the Strong CP Problem. It would freeze the axion field at the value $a/F_a = - \theta_h$ and prevent it from adjusting to cancel $\theta_{QCD}$. In fact, the effective strong CP phase would be $\overline{\theta} \cong - N_{QCD} \theta_h + \theta_{QCD}$.

This difficulty would be avoided, however, if the potential $V_h(a)$ turned off again before the temperature fell to $\Lambda_{QCD}$. 
This can happen as follows. It is well-known that in QCD if any quark were exactly massless the Strong CP Problem would be solved, since the Strong CP phase $\overline{\theta}$ could be absorbed into a phase redefinition of the massless quark field. For the same reason, a massless quark would make physics invariant under a shift $a(x^{\mu}) \rightarrow a(x^{\mu}) + \theta$,
meaning that the potential $V_{QCD}(a)$ would be flat. (This is why the axion mass comes out proportional to the squareroot of the mass of the lightest quark \cite{KimRMP10}.) In an analogous way, if some $G_h$-quark had a zero mass, then the potential $V_h(a)$ would be flat. Suppose, therefore, that in addition to $Q$, which has non-zero Peccei-Quinn charge, we assume there exists another $G_h$-quark $Q'$ that has vanishing Peccei-Quinn charge. The mass of $Q'$, therefore, does not come from the vacuum expectation value of $\Phi$ (whose phase is $a/F_a$), but from the vacuum expectation value of some other scalar field $X$ that does not transform under the Peccei-Quinn symmetry. Suppose that there is a phase transition at some critical temperature $T_c$, where $\Lambda_{QCD} \ll T_c < \Lambda_h$, such that $\langle X \rangle \neq 0$ above $T_c$ but $\langle X \rangle = 0$ below $T_c$.
(Such inverted phase transitions are possible and have been studied in the past \cite{inverted-phase-transition}.)
What happens in this case is that when $T$ falls below $T_c$, the quark $Q'$ becomes exactly massless, and $V_h(a)$ becomes flat, i.e. turns off, and the axion dynamics and axion mass would be controlled by $V_{QCD} (a)$ alone, allowing the standard axion solution of the Strong CP Problem to work. (It should be noted that a massless $Q'$ would be confined by the non-abelian $G_h$ force into $G_h$ ``hadrons" whose mass was of order $\Lambda_h$, and thus too large to have been observed.)

One sees that we have paid some price in complexity to solve the axion domain-wall problem, for we have now introduced not only the $G_h$ interaction and $G_h$-quark $Q$, but also the fields $Q'$ and $X$, and any further fields required to give $X$ an inverted phase transition.

Let us now return to the question of how much the ``decay axions" contribute to the dark matter density. In our scenario, domain walls form when the potential $V_h$ turns on at a temperature of order $\Lambda_h$. These walls get chopped up by the Vilenkin-Everett
mechanism, and the resulting finite loops of string and patches of domain wall radiate their energy into axions. However, when $T$ falls to $T_c$, and $V_h$ turns off, these decay axions
become massless. At that point, as we will now see, their contribution to the energy density of the universe is very small compared to that of thermal energy in other massless particles and remains small.

Let the temperature at which the axion field begins to oscillate due to
$V_h$ be called $T_*$, with $T_* \sim \Lambda_h$. One therefore has, roughly, that $m_a(T_*) \sim H(T_*) \sim T_*^2/M_P$. First, let us consider the axions in these coherent oscillations.\footnote{ Above $T^*$, the axion mass is suppressed exponentially \cite{BHK08}.
}
 For these coherent axions one has
$\rho_a(T_*) \sim m_a(T_*) n_a(T_*) \sim
m_a^2(T_*) F_a^2$, so that $n_a(T_*) \sim m_a(T_*) F_a^2 \sim T_*^2 F_a^2/M_P$.
These coherent axions have typical wavelength $\lambda (T_*) \sim m_a^{-1} (T_*) \sim M_P/T_*^2$.
When the temperature falls to $T_c$, the (now massless) axions have number density and typical
wavelength given by $n_a(T_c) \sim (T_*^2 F_a^2/M_P)(T_c/T_*)^3 =
(F_a^2 T_c^3)/(M_P T_*)$ and $\lambda (T_c) \sim (M_P/T_*^2)(T_*/T_c) =
M_P/(T_c T_*)$. The energy density in these massless axions is thus $\rho_a (T_c)
\sim n_a(T_c) \omega (T_c) \sim (F_a/M_P)^2 T^4 \ll T^4$, {\it i.e.} much less than the
thermal energy in other massless particles.

%%%%%%%%%%%%%%%%%%%%%%%%%%%%%%%%%%%%%%%%%%%%%%%%%%%%%%%%%%%%
%\begin{figure}[!t]
%\begin{center}
%\includegraphics[width=0.35\linewidth]{figQCDwall.eps}
%\end{center}
%\caption{QCD vacuum angles. Around the string, the minimum of the hidden sector potential is at $\theta =\frac{\pi}{2}$ and the minimum of the QCD potential is at $\theta =0$.}\label{fig:thetaQCD}
%\end{figure}
%%%%%%%%%%%%%%%%%%%%%%%%%%%%%%%%%%%%%%%%%%%%%%%%%%%%%%%%%%%%

The amplitude of the coherent axion oscillations evolves in time in the following manner. When the oscillations due to $V_h$ begin (at
$T = T_*$), they have amplitude of order $F_a$ and the axion mass is $m_a(T_*) \sim T_*^2/M_P$. As the axion mass due to $G_h$ instantons turns on adiabatically to its full value $\overline{m}_a \sim
\Lambda_h^2/F_a$, the amplitude of the coherent oscillations is reduced by a factor
$(m_a(T_*)/\overline{m}_a)^{1/2} \sim (T_*/\Lambda_h) \sqrt{F_a/M_P}$. (This is because the number density of coherent axions, which is proportional to $ma^2$ is an adiabatic invariant \cite{coherent-osc}.)  As the temperature then drops to $T_c$, the amplitude of axion oscillations is reduced by a further factor
$(T_c/T_*)^{3/2}$, due to the expansion of the universe. Finally, the axion mass starts to turn off adiabatically when $T \sim T_c$, and axion oscillations cease at a temperature $T'_* \sim T_c$ where $m_a(T'_*) \sim
T^{'2}_*/M_P$. As the axion mass changes from $\overline{m}_a$ to $m_a(T'_*)$, the amplitude of axion oscillations grows by a factor $(\overline{m}_a/m_a(T'_*))^{1/2} \sim
(\Lambda_h/T'_*) \sqrt{M_P/F_a}$ (again, because the number density is an adiabatic invariant \cite{coherent-osc}). Altogether, then, multiplying these factors, one finds that the amplitude of the axion oscillations produced by $V_h$ at the time when they cease at $T'_*$ is of order
$F_a [(T_*/\Lambda_h) \sqrt{F_a/M_P}] [T_c/T_*]^{3/2}
[(\Lambda_h/T'_*) \sqrt{ M_P/F_a}]
= F_a \frac{T_c^{3/2}}{T_*^{1/2} T'_*}$. Since $T_* \sim \Lambda_h$ and $T'_* \sim T_c$.
The amplitude of the coherent axion oscillations due to $V_h$ when they cease
is of order $F_a (T_c/\Lambda_h)^{1/2} \ll F_a$. In other words, the amplitude of the coherent oscillations in the angle $a/F_a$ due to the potential $V_h$ become small as that potential turns on and then remain small when it turns off again. As a consequence, the variations in $a/F_a$ these oscillations cause will be very small compared to the misalignment of $a/F_a$ when $V_{QCD}$ turns on, and have a negligible effect on
the present axion energy density.

%%%%%%%%%%%%%%%%%%%%%%%%%%%%%%%%%%%%%%%%%%%%%%%%%%%%%%%%%%%%
%%\begin{figure}[!t]
%%\begin{center}
%%\includegraphics[width=0.35/linewidth]{figQCDwall.eps}
%%\end{center}
%%\caption{QCD vacuum angles. Around the string, the minimum of the hidden sector potential %%is at $\theta =\frac{\pi}{2}$ and the minimum of the QCD potential is at $\theta %%=0$.}\label{fig:thetaQCD}
%%\end{figure}
%%%%%%%%%%%%%%%%%%%%%%%%%%%%%%%%%%%%%%%%%%%%%%%%%%%%%%%%%%%%

One sees, then, that the coherent axions produced
when $V_h$ turns on have negligible effect on the present dark matter energy density.

The analysis of the axions produced when strings and walls decay is very similar.
When the domain walls form at $T_*$, the horizon length is
$\ell_* = H^{-1} (T_*) \sim M_P/T_*^2$. There is typically one horizon-length string per
Hubble volume. A horizon-length string has mass
$m_{str} (T_*) \sim F_a^2 \ell_*$, and therefore the energy density in strings is of order
$\rho_{str}(T_*) \sim F_a^2 \ell_*^{-2}$. This energy is radiated into axions of
typical wavelength $\ell_*$. Thus the number density of these ``decay axions" is of order
$n_a(T_*) \sim F_a^2 \ell_*^{-1} \sim T_*^2 F_a^2/M_P$. One sees that this is of the same order as the number density of coherent axions. Moreover, their typical wavelengths are the same. Thus the analysis we made of the coherent axions applies to the decay axions as well.
By the time the temperature falls to $\Lambda_{QCD}$, all the energy in axions coming from the strings, walls, and coherent oscillations that were due to $V_h$ have become negligible. And no new strings form when $T \sim \Lambda_{QCD}$, so that the axion dark matter present today is virtually entirely due to the coherent oscillations coming from
$V_{QCD}$. In effect, then, we have made $\adec$ in Eq. (2) equal to zero.

Up to this point we have been discussing a scenario where there is a new confining force that generates a potential $V_h$ and gives the axion a mass when the temperature is high compared to the QCD scale, thereby allowing the Vilenkin-Everett mechanism to get rid of all the strings before the QCD instanton effects turn on. However, we can see that the
potential $V_h$ could be generated in a different way, without a new confining force.
Suppose, for example, that (just as in the previous discussion) there are fields $\Phi$ and $X$, where $\Phi$ breaks the Peccei-Quinn symmetry spontaneously at a scale $F_a$ and $X$ has an inverted phase transition at
some scale $T_c \ll F_a$. And suppose that these fields have a tree-level potential
$V(\Phi, X) = \frac{1}{4} \lambda ( | \Phi|^2 - F_a^2)^2 + \frac{1}{2} g X^2 (\Phi + \Phi^*) + V(X)$.

When $T \ll f_a$, but $> T_c$, one has $\langle X \rangle \neq 0$, and the axion has a potential $V_h (a) = \frac{1}{2} g \langle X \rangle^2 (F_a e^{ia/F_a} + F_a e^{-ia/F_a})
= g \langle X \rangle^2 F_a \cos(a/F_a) \cong \frac{1}{2} (g \langle X \rangle^2/ F_a)
a^2$. This gives a unique minimum ({\it i.e. $N = 1$}). Thus the walls that are produced will lead to the destruction of the string-wall system by the Vilenkin-Everett mechanism.
When $T$ falls to $T_c$, the vacuum expectation value of $X$ disappears and the tree-level $V_h$ disappears. The scenario is thus very similar to the one discussed before. One difference is that there are one-loop diagrams involving the coupling $g$ that give a small
mass to the axion for low temperatures. This potential has the form $V'_h(a)\sim
\frac{1}{16 \pi^2} g^2 (\Phi^2 + \Phi^{*2}) =
\frac{1}{16 \pi^2} g^2 a^2$. One must have $g$ small enough so that this loop-induced
potential does not interfere with the axion solution to the Strong CP Problem, but large enough that the domain walls produced by $V_h$ can eliminate the string-wall system by the Vilenkin-Everett mechanism. The energy/length of a string is of order $F_a^2$. The surface tension of the walls produced by $V_h$ is of order $F_a^2 m_a \sim F_a^2 (g \langle X \rangle^2/F_a)^{1/2}$. The condition that the walls are chopped up by the Vilenkin-Everett mechanism by the time temperature of the universe is $T_c$ is roughly that $(g \langle X \rangle^2/F_a)^{1/2} >
T_c^2/M_P$. If we take $T_c \sim \langle X \rangle$, then this gives roughly
$g > \langle X \rangle^2 F_a/M_P^2$.  On the other hand, in order for the potential $V'_h(a)$ not to prevent the PQ solution to the String CP Problem, one must have
$g^2 < \overline{\theta} (\Lambda_{QCD}^4/f_a^2)$. Combining these two limits on $g$, one
obtains $\langle X \rangle < \overline{\theta}^{1/4} M_P \Lambda_{QCD} f_a^{-1} N_{QCD}^{-1/2}$ (recalling that $f_a = F_a/N_{CQD}$). If we take
$f_a \sim 10^{12}$ GeV, these conditions can be satisfied with $\langle X \rangle \sim 1$ TeV, and $g \sim 10^{-18}$ GeV, which suggests that the term $g X^2 \Phi$ comes from a Planck-suppressed higher-dimension operator.

This last model is not meant as a fully realistic one, but only to illustrate that the $V_h$ required to eliminate strings by our mechanism can arise in another way than from a new confining force.

%%%%%%%%%%%%%%%%%%%%%%%%%%%%%%%%%%%%%%%%%%%%%%%%%%%%%%%%%%%%%%%%%%%%%%%%%%%%%%%%%%%%%%%%%%%%%

%\section{Conclusion}
\noindent{\it Conclusion}--
In typical axion models, axion domain walls form when the temperature reaches the QCD scale. If $N >1$ walls are attached to each axion string, then a persistent
string- wall network forms that overcloses the universe. But even if each string bounds just one wall, so that axion walls can be eliminated by the Vilenkin-Everett mechanism, the axion particles so radiated would augment the axion dark matter density, which would  tightens constraints on models of axion dark matter.

We have proposed a mechanism by which axion strings can be eliminated long before $T$ reaches the QCD scale, thus preventing the formation of problematic axion domain walls at that scale. In this scenario, an axion potential $V_h$ generated by some non-QCD interaction turns on when $T \gg \Lambda_{QCD}$. This potential produces axion domain walls with $N=1$ that eliminate the strings by the Vilenkin-Everett mechanism. Consequently, no axion domain walls are produced at the QCD phase transition, as they have no strings to attach to. This scenario requires that $V_h(a)$ turn off again at some temperature $T_c > \Lambda_{QCD}$, as otherwise $V_h(a)$ would interfere with the axion solution of the Strong CP Problem \cite{StrongCP}.

Turning off $V_h(a)$ requires introducing a sector of fields that undergoes an inverted phase transition of the type discussed in \cite{inverted-phase-transition}. This, of course, involves some complication of the model.

Some issues bear further investigation. It is worth doing a more detailed calculation of the density of axions radiated from the $V_h$ walls to confirm that it is negligible. It would be useful to construct more detailed models, including specific dynamics for the required inverted phase transition.

%\end{document}

\section*{Acknowledgments}
SMB is supported in part by DOE grant No. DE-FG02-12ER41808, and JEK is supported in part by the National Research Foundation (NRF) grant funded by the Korean Government (MEST) (No. 2005-0093841) and  by the Institute of Basic Sciences (IBS-R017-D1-2014-a00).

\end{document}